# Prefix based Chaining Scheme for Streaming Popular Videos using Proxy servers in VoD

M Dakshayini,
Research Scholar, Dr. MGR University. Working with Dept. of ISE, BMSCE,
Member, Multimedia Research Group, Research Centre, DSI,
Bangalore, India.
shantha_dakshu@yahoo.co.in

Dr T R GopalaKrishnan Nair
Director, Research and Industry Incubation Centre, DSI,
Bangalore, India
trgnair@yahoo.com

*Abstract*—Streaming high quality videos consumes significantly large amount of network resources. In this context request-to-service delay, network traffic, congestion and server overloading are the main parameters to be considered in video streaming over the internet that effect the quality of service (QoS). In this paper, we propose an efficient architecture as a cluster of proxy servers and clients that uses a peer-to-peer (P2P) approach to cooperatively stream the video using chaining technique. We consider the following two key issues in the proposed architecture (1) Prefix caching technique to accommodate more number of videos close to client (2) Cooperative client and proxy chaining to achieve the network efficiency. Our simulation results shows that the proposed approach yields a prefix caching close to the optimal solution minimizing WAN bandwidth usage on server-proxy path by utilizing the proxy-client and client-client path bandwidth, which is much cheaper than the expensive server –proxy path bandwidth, server load, and client rejection ratio significantly using chaining.

## I. INTRODUCTION

Generally streaming any multimedia object like high quality video consumes a significantly large amount of network resources. So request-to-service delay, network traffic, congestion and server overloading are the main parameters to be considered in video streaming over the internet that affect the quality of service (QoS). So providing video-on-demand (VoD) service over the internet in a scalable way is a challenging problem. The difficulty is twofold first; it is not a trivial task to stream video on an end-to-end basis because of a video's high bandwidth requirement and long duration. Second, scalability issues arise when attempting to service a large number of clients. In particular, a popular video generally attract a large number of users that issues requests asynchronously [2]. There are many VoD schemes proposed to address this problem: batching, patching, periodical broadcasting and prefix caching and chaining.

In the batching scheme [5,8 &10], the server batches the requests for the same video together if their arrival times are close, and serve them by one multicast channel. In the patching scheme [2], the server sends the entire video clip to the first client. Later clients can join the existing multicast channel, and at the same time each of them requires a unicast channel to deliver the missing part of the video.

Periodical broadcasting [12] is another innovative technique. In this approach, popular videos are partitioned into a series of segments and these segments are continually broadcasted on several dedicated channels. Before clients start playing videos, they usually have to wait for a time length equivalent to the first segment. Therefore, only near VoD service is provided.

Proxy caching [1, 4 & 9] is also a promising scheme to alleviate the bandwidth consumption issue. In this approach, there exists a proxy between a central server and client clouds. Partial video (or entire video) files are stored in proxies and the rest are stored in the central server. Proxies send cached videos to clients, and request the remaining from servers on behalf of clients. Recent works investigate the advantages of connected proxy servers within the same intranet [3, 4 and 8].

## II. RELATED WORK

In this section we briefly discuss the previous work. Tay and pang has proposed an algorithm in [7] called GWQ (Global waiting queue) to reduce the initial startup delay by sharing the videos in a distributed loosely coupled VoD system. They have replicated the videos evenly in all the servers, for which the storage capacity of individual proxy server should be very large to store all the videos. In [11] Sonia Gonzalez, Navarro, Zapata proposed an algorithm to maintain a small initial start up delay using less storage capacity servers by allowing partial replication of the videos. They store the locally requested videos in each server. We differ by caching the partial prefix-I at proxy and prefix-II at tracker in proportion to popularity there by utilizing the proxy server and tracker storage space more efficiently. In [3] authors have proposed an hybrid algorithm for chaining, but they do not discuss about the scenario of client failure. LEMP in [6] also have proposed a client chaining mechanism and a solution for handling client failure situation involving too many messages which increases the waiting time for playback start to $t_w$. Another approach to reduce the aggregated transmission cost has been discussed in [12] by caching the prefix and prefix of suffix at proxy and client respectively. But they have not considered chaining. [5 and 8] proposes a batching technique, which increases the client initial waiting time. Edward Mingjun Yan and Tiko kameda in [10] proposes a broadcasting technique which requires huge amount of storage capacity and sufficient bandwidth. Yang Guo in [2] has suggested an architecture to stream the video using patching technique. Hyunjoo and Heon in [13] have proposed another chaining scheme with VCR operations. But they do stream the video data from main server and considered a constant threshold value, due to which more number of clients may not be able to share the same chain. And WAN bandwidth usage on server-proxy path may be comparatively high.

In this paper, we propose an efficient load sharing algorithm and a new VoD architecture for distributed VoD system. This architecture consists of a centralized Main multimedia server [MMS] which is connected to a set of trackers [TR]. Each tracker is in turn connected to a group of proxy servers [PS] and these proxy servers are assumed to be interconnected using ring pattern. To each of this PS a set of clients are connected. And all these clients are connected among themselves. This arrangement is called as Local Proxy Servers Group[LPSG($L_p$)]. Each of such LPSG, which is connected to MMS, is in turn connected to its left and right neighboring LPSG in a ring fashion through its tracker. And an efficient prefix caching based chaining (PC+Chaining) scheme using proxy servers to achieve the higher network efficiency.

The organization of rest of the paper is as follows: Section 3 analyzes various parameters used in the problem. In section 4 we present a proposed approach and algorithm in detail, Section 5 describes Simulation model and section 6 describes the PC+Chaining scheme in detail, Section 7 presents the performance evaluation, Finally, in section 8, we conclude the paper and refer to further work.

### III. MODEL OF THE PROBLEM

Let $N$ be a stochastic variable representing the group of videos and it may take the different values (videos) for $V_i$ ($i=1,2..N$). And the probability of asking for the video $V_i$ is $p(V_i)$, let the set of values $p(V_i)$ be the probability mass function. Since the variable must take one of the values, it follows that $\sum_{i=1}^{N} p(v_i) = 1$.

So the estimation of the probability of requesting $V_i$ video, is

$p(V_i) = \dfrac{n_i}{I}$.

Where $I$ is the total number of observations and $n_i$ is the number of requests for video $V_i$. We assume that client's requests arrive according to Poisson process with the arrival rate $\lambda$. Let $S_i$ be the size (duration in minutes) of $i^{th}$ video($i=1..N$) with mean arrival rates $\lambda_1 \ldots \lambda_N$ respectively that are being streamed to the users using $M$ proxy servers (PSs) of $J$ LPSGs ($L_p$ $p=1..J$). Each TR and $PS_q$ ($q=1..M$), has a caching buffer large enough to cache total $P$ and $B$ minutes of $H$ and $K$ number of video prefixes respectively. The complete video is divided into three parts, first $W_1$ minutes of each video $V_i$ is referred to as prefix-1$(pref-1)_i$ of $V_i$ and is cached in any one of the proxy servers of the group only once. and next $W_2$ minutes of video $V_i$ is referred to as prefix-2$(pref-2)_i$ of $V_i$ is cached at TR of $L_p$.

i.e. $P = \sum_{i=1}^{H}(pref\text{-}2)_i$ and $B = \sum_{i=1}^{K}(pref\text{-}1)_i$

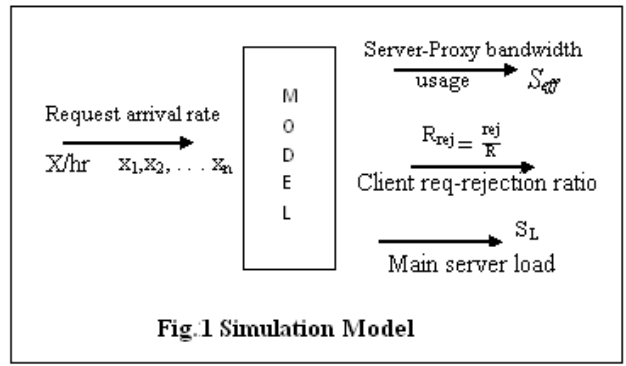

Fig.1 Simulation Model

Based on the frequency of user requests to any video, the popularity of the videos and size of *(pref-1)* and *(pref-2)* to be cached at PS and TR respectively is determined. And the size (W) of *(pref-1)* and *(pref-2)* for $i^{th}$ video is determined as.

$W(pref-1)_i = x_i \times S_i$ where $0 < \chi_i < 1$

$W(pref-2)_i = x_i \times (S_i - (pref\text{-}1)_i)$ where $0 < \chi_i < 1$

Where $x_i$ is the probability of occurrence of user requests with frequency for video $i$ from last $t$ minutes. This arrangement caches maximum portion of most frequently requesting videos at $L_p$. So most of the requests are served immediately from Lp itself, which reduces the network usage on server-proxy path significantly and makes the length of the queue $Ql$ almost negligeble. Let $b_i$ be the available bandwidth for $V_i$ between the proxy and main Server. After requesting for a video $V_i$ at $PS_q$, the WAN bandwidth required on server-proxy path for the video $V_i$ may be

$bw_i^{S-P} = bw(S - (pref\text{-}1) - (pref-2))_i^{S-P}$

where $i=1..N$, and $bw_i^{S-P}$ is the WAN bandwidth usage required for $i^{th}$ video on server to proxy path. This $bw_i^{S-P}$ depends on the amount of $V_i$ (S-(pref-1)-(pref-2)) to be streamed from main server to the proxy. So the aggregate server-proxy bandwidth usage would be

$$WAN_{bw}^{S-P} = \sum_{i=1}^{Q} bw(S-(pref-1)(pref-2))_i^{S-P}$$

$bw()$ is a non linear function of $Sz[(pref-1) \& (pref-2)]$. And another output stochastic variable $R_{rej}$ is the request rejection ratio, which is the ratio of the number of requests rejected ($N_{rej}$) to total number of requests arrived at the system (R), which is inversely proportional to system efficiency $S_{eff}$.

That is $S_{eff} = \dfrac{1}{R_{rej}}$ where $R_{rej} = \dfrac{N_{rej}}{R}$ and

$S_{eff} = \dfrac{Q}{R}$ is the ratio of number of requests served (Q) to total number of requests arrived (R) at the system. The optimization problem to maximize $S_{eff}$, thereby minimizing the client rejection ratio $R_{rej}$ at the PS, and average WAN bandwidth usage $WAN_{bw}^{S-P}$ is.

Maximize System Efficiency $S_{eff} = \dfrac{Q}{R}$

Minimize Avg WAN Bandwidth usage

$$WAN_{bw}^{S-P} = \sum_{i=1}^{Q} bw(S-(pref-1)(pref-2))_i^{S-P}$$

$$y = \dfrac{1}{Q} \sum_{i=1}^{Q} (Wt_i)$$

Subject to

$$B = \sum_{i=1}^{K}(pref-1)_i, \quad P = \sum_{i=1}^{H}(pref-2)_i$$

$$W(pref-1) \& W(pref-2) > 0$$

IV. PROPOSED ARCHITECTURE AND ALGORITHM

A. Proposed Architecture

The proposed VoD architecture is as shown in Fig.2. This architecture consists of a MMS, which is connected to a group of trackers (TRs), As shown in Fig.3, each TR has various modules like Interaction Module ($IM_{TR}$) – Interacts with the PS and MMS. Service Manager ($SM_{TR}$) – Handles the requests from the PS. Database – Stores the complete details of presence and size of (pref-1) of videos at all the PSs. Video distributing Manager(VDM) – Responsible for deciding the videos, and sizes of (pref-1), (pref-2) of videos to be cached. Also handles the distribution and management of these videos to group of PSs, based on video's popularity.

And each TR is in turn connected to a set of PSs. These PSs are connected among themselves in a ring fashion. Each PS has various modules like Interaction Module of PS ($IM_{PS}$)– Interacts with the client and TR. Service Manager of PS ($SM_{PS}$)–Handles the requests from the user, Client Manager (CM) – Observes and updates the popularity of videos at PS as well as at TR as shown in Fig.3. And also to each of these PSs a large number of users are connected [LPSG]. Each proxy is called as a parent proxy to its clients. All these LPSGs are interconnected through their TR in a ring pattern. The PS caches the (pref-1) of videos distributed by VDM, and streams this cached portion of video to the client upon the request through LAN using its less expensive bandwidth. We assume that,

1. The TR is also a PS with high computational power and large storage compared to other proxy servers, to which clients are connected. It has various modules, using which it coordinates and maintains a database that contains the information of the presence of videos, and also size of (pref-1) and (pref-2) of video in each PS and TR respectively

2. Proxies and their clients are closely located with relatively low communication cost[1]. The Main server in which all the videos completely stored is placed far away from LPSG, which involves high cost remote communication.

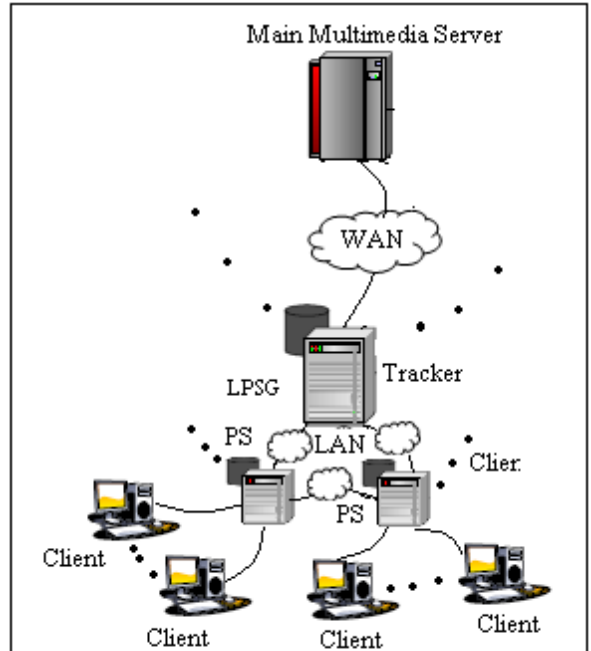

Fig. 2 Architecture of Cluster of Proxy Servers

3. The MS, the TR and the PSs of LPSG are assumed to be interconnected through high capacity optic fiber cables.

In the beginning, all the $N_v$ videos are stored in the MMS. The distribution of the selected N of $N_v$ videos among M PSs in an LPSG is done by Video Distribution Manager of TR as follows. First, all the N videos are arranged with respect to their popularity at jth LPSG. The popularity of a video is defined as the frequency of requests to this video per threshold time t by the clients. Here, we assume that the frequency of requests to a video follows Zipf law of distribution.

### B. Proposed Algorithm

```
1. When a request from C_k for video V_i arrives at time t to
   particular proxy PS
2. Do the following
3. If (V_i is present at PS)
4.    Check the SCL
5.    If ((IS-STREAMING_{V_i} is TRUE) && ((t_{C_j} - t_{C_k}) ≤ (prif-1)_{V_i}))
6.                 where j=1..d
7.       req-hdlr of PS sends msg with LAC to C_k and updates SCL
8.    else ser-mgr starts the new stream to C_k and add C_k to SCL
9. else pass the Req to TR, TR checks in Lp
10. If (V_i is present at some PS of Lp)
11.    TR initiates the new stream to C_k thru PS & ser-mgr
                updates the SCL
12. else TR checks at NBR[Lp]
13. If (V_i is present at some PS of NBR[Lp])
14.    TR initiates the new stream to C_k thru PS
                & ser-mgr update the SCL
15. else TR decides to download the V_i from MMS
16.    Start new stream to C_k,& comute (pref-1)_{V_i}, (pref-2)_{V_i}
17. If ( (pref-1)_{V_i}=(x_i × S_i) & (pref-2)_{V_i}=(x_i ×(S_i-(pref-1)))) can
                be cached
18.    Cache (pref-1)_{V_i} & (pref-2)_{V_i} at PS & TR respectively and
                update SCL
19. else try to make room for (pref-1)&(pref-2) at Lp using
                some buffer allocation algorithm
```

Whenever a client $C_k$ request for the video $V_i$ at *PS*, *request-handler* (req-handler) checks whether *IS-STREAMING* flag of that video is true and the *arrival time difference* of the new client and the most recent client of the existing chain of $V_i$ is below the threshold $W(pref-1)$ then the *service-manager (service-mgr)* adds $C_k$ to the existing client chain of $V_i$ and instructs it to get streamed from any one of the last *d* clients of the chain. Then by adding $C_k$ to the client list of *Vi*, *service-mgr* updates the *Streaming Clients List (SCL)*, which is the list with complete details of list of videos being streamed from that PS and the corresponding chain of clients of that video [line 3-7 of proposed algorithm]. Otherwise *PS* starts the new stream to $C_k$ and new chain to $V_i$ and *SCL* is updated by creating a new entry for *Vi* [line 8]. If it is not present in its cache, the $IM_{PSq}$ forwards the request to its parent *TR*, *VDM* at *TR* searches its database using perfect hashing to see whether it is present in any of the *PSs* in that $L_p$. If it finds starts streaming from that *PS* to $C_k$ and updates the *SCL* accordingly [line 9,10 &11].

Otherwise request is passed to *TR* of *[NBR[Lp]]*. If *Vi* is found there it will be streamed from that *PS* to $C_k$ and *SCL* is updated accordingly by *service-manager* [line 12,13 & 14]. If *Vi* is not found at *NBR[Lp]* also then the *Vi* is downloaded from the MMS and is streamed to *Ck*. While streaming the *W(pref-1)* and *W(pref-2)* of $V_i$ is calculated as

$$W(pref-1)_i = x_i \times S_i$$ where $0 < \chi_i < 1$ and

$$W(pref-2)_i = x_i \times (S_i - (pref-1)_i)$$

where $0 < \chi_i < 1$

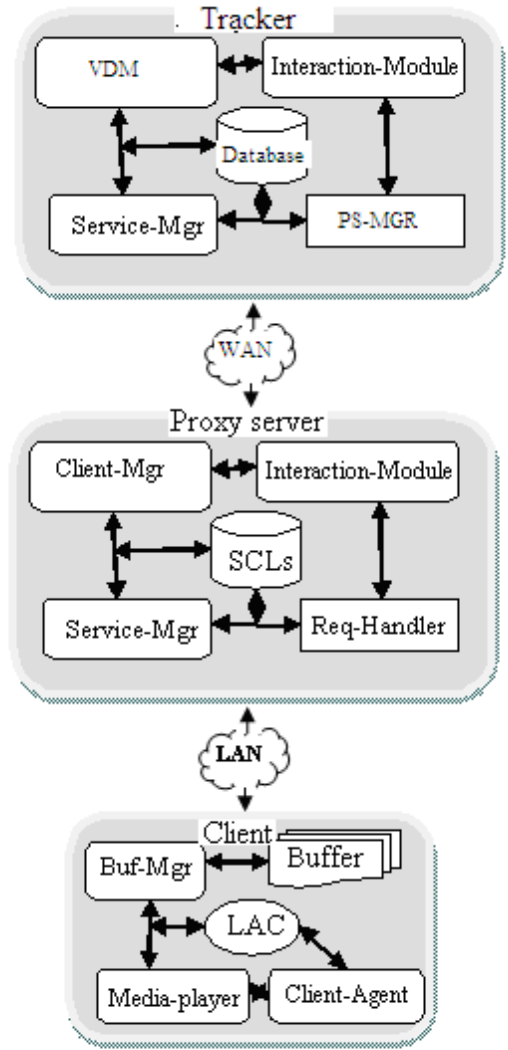

Fig. 3 Tracker Proxy and Client Modules

and cached at *PS* and *TR* respectively, if sufficient space is available. Otherwise an appropriate buffer allocation algorithm is executed to make space to cache these prefixes of $V_i$ according to popularity [line 15,16 17 and 18]. So most of the requests are served immediately from the client, who is already being served or from $L_p$ by sharing the videos present among the *PSs*, which reduces the *client rejection-request ratio $R_{rej}$, load at the server $S_L$, and increases the video hit ratio VHR.*

## V. SIMULATION MODEL

In our simulation model we have a single *MMS* and a group of 6 *TRs*. All these *TRs* are interconnected among themselves in a ring fashion. Each of these *TR* is in turn connected to a set of 6 *PSs*. These *PSs* are again interconnected among themselves in a ring fashion. To each of these PS a set of 25 clients are connected and all these clients are interconnected among themselves. We use the video hit ratio (*VHR*), the average client waiting time $y$ to measure the performance of our proposed approach more correctly. In addition we also use the *WAN* bandwidth usage on *MMS-PS* path and probability of accessing the main server as the performance metrics.

Table .1
Simulation Model

| Notation | System Parameters | Default Values |
|---|---|---|
| $S$ | Video Size | 2-3hrs, 200units/hr |
| $C_{MMS}$ | Cache Size(*MMS*) | 1000 |
| $C_{TR}$ | Cache Size(*TR*) | 400(40%) |
| $C_{PS}$ | Cache Size(*PS*) | 200(20%) |
| $\lambda$ | Mean request arrival rate | 44 reqs/hr |

We assume that the request distribution of the videos follows a zipf-like distribution. The user request rate at each *PS* is 35-50 requests per minutes. The ratio of cache sizes at different elements like *MMS*, *TR* and *PS* is set to $C_{MMS} : C_{TR} : C_{PS} = 10:4:2$ and transmission delay between the proxy and the client, proxy to proxy and *TR* to *PS* as 100ms, transmission delay between the main server and the proxy as 1200ms, transmission delay between tracker to tracker 300ms, the size of the cached *[(pref-1)+(pref-2)]* video as 280MB to 1120MB(25min – 1hr) in proportion to its popularity.

## VI. PC+CHAINING

Our proposed scheme PC+Chaining is an efficient streaming technique that combines the advantages of proxy prefix caching and both client-server and peer-to-peer approaches to cooperatively stream the video using chaining. The main goal of PC+Chaining is to make each client act as a server while it receives the video, so that the available memory and bandwidth of the clients can be utilized more efficiently. The un-scalability of traditional client-server unicast VoD service lies in the fact that the server is the only contributor and can thus become flooded by a large number of clients submissively requesting the service. In the client-server service model, a client sets up a direct connection with the server to receive the video. In this case an amount of WAN bandwidth requirement on server-proxy path equal to the playback rate is consumed along the route. As the number of requests increases, the bandwidth at the server and in the network is consumed so network becomes congested and the incoming requests must eventually be rejected [2]. In contrast, we propose two schemes to address these issues 1) Local group of interconnected proxies' and clients architecture with prefix caching technique and load sharing among the proxies of the group reduces the frequent access to MMS which in turn reduces the amount of bandwidth consumption between client and main server. 2) PC+Chaining, where the clients not only receive the requested stream, but also contribute to the overall VoD service by forwarding the stream to other clients, whose request arrives within the threshold time of *W(pref-1)*. In PC+Chaining all the clients are treated as potential server points. When there is a request for a video $V_i$ from the client $C_k$ at particular proxy $PS_q$, if the requested video $V_i$ is present at $PS_q$, then the service will be provided to $C_k$ in the following stages

### C. Client admission phase

When the request arrives at $PS_q$, the *request-handler* (req-handler) of that proxy checks for the presence of the video at $PS_q$. If it is present then it checks the flag *IS-STREAMING* of the video $V_i$. If it is not true indicates that, there are no clients existing having streamed with the same video object $V_i$. Then the *req-handler* informs the *service-manager (service-mgr)* to provide the streaming of $V_i$ to $C_k$. So the *service-mgr* starts new stream and updates the *streaming clients list (SCL)* by adding a new entry for the video $V_i$ along with its (*pref-1*) size. The format of each entry of *SCL* is <*video id - sz(pref-1)- list of clients being streamed with $V_i$*> If the flag *IS-STREAMING* is true, indicating that there is already a chain of clients being streamed with the same video. Then the *req-handler* looks into the corresponding entry for $V_i$ in *SCL* to check whether the arrival time difference of the new client and the most recent client of the existing chain of $V_i$ is below the threshold *W(pref-1)*, if so $C_k$ is added to the existing chain of $V_i$ and sends the *sub list of d applicant clients (LAC)* to $C_k$, along with the *msg* to get streamed from the last client of the list *LAC*. Also sends the msg to all d clients of *LAC*, so that the *client agent* of these clients sends ready signal to $C_k$. Then $C_k$ sends start signal generally to the most recent client being streamed with $V_i$, and starts getting the stream. This client becomes the *parent client(PC)* of $C_k$. In case any problem with this client, $C_k$ can request the other clients.

At the end of this client admission phase, *SCL* will have the complete details of list of videos being streamed and the corresponding chain of clients of that video along with the prefix size. Also each client knows its applicant ancestors to change its *parent client (PC)* in case of *PC* failure.

### D. Streaming phase

We assume that the client has sufficient buffer space and network bandwidth to accommodate the (pref-1) of $V_i$. when the client $C_k$ requested for $V_i$ starts getting the stream from the last client of *LAC*, the *buffer manager (buf-mgr)* starts buffering the video clips received from its parent client. And *media-player* starts playing it from the buffer.

### E. Closing Phase

Any client $C_k$ in the chain may wants to close the connection in the following two cases

case1: Once the streaming of the complete video is finished. The *client agent* closes the connection with its *parent client (PC)* by sending *close* signal to it, but checks whether this client is *PC* for any other clients in the chain before sending *close* signal to its *parent proxy server(PPS)*. If so, it waits until the streaming of the complete video to its *child client (CC)* finishes and then sends the *close* signal to its *PPS* and then closes the connection. The *service-mgr* updates the case2: In the middle of the streaming, if the $C_k$ wants to stop watching the video, first *client-agent* of $C_k$ checks whether it is a parent for any other client in the chain. If so it checks $(t_{pc}(C_k) - t_{cc}(C_k))_{Vi} < W(pref-1)_{Vi}$ where $t_{pc}(C_k)$ is the arrival time of *PC* of $C_k$ and $t_{cc}(C_k)$ is the arrival time of *CC* of $C_k$. If yes it sends the *close* flag to its *PC, CC* and to its *PPS* indicating that it is going to be closed and also gives the details of its *CC* to its *PC*. So the *CC* of $C_k$ can continue getting the stream from *PC* of $C_k$ and informs *PPS* about the change of its *PC*. Then *service-mgr* of *PPS* updates the *SCL* by removing the presence of $C_k$ from the client list of $V_i$

### F. Client Failure recovery

At any time it is possible for clients to fail, withdraw or operate in a lossy network environment [3]. A client's failure may break the chain causing all its descendant clients loosing the streaming data. Because a failed client can no longer stream the video data to its descendant clients, so the buffered video data of its *CC* may fall below the threshold, thus the *CC* can detect the failure of its *PC*. So after a failure, its *CC* can detect the failure and notify the *PC* of the failed client. Also notify the *PPS* about the failure of its *PC*. So the *service–mgr* of *PPS* can update the SCL accordingly. Fig.4 shows an example of client chain for video56 and for video 14. Also shows the *SCL* with complete details of the *video-id*, *sz(pref-1)* and the *list of clients* in the chain.

Fig.5 shows a situation when one of the clients *C4* of the chain for the video 14 fails C5 which is the *CC* of *C4*, detects the failure of *C4*, when it does not receive the video data even after *g* seconds. C5 then notifies both the *PPS* and *C3* which is *PC* of *C4* along with its address and the details of its streaming like where it was stopped. Then *PC* of *C4* stops streaming to *C4* starts streaming to C5 and becomes the *PC* of *C5*, shown as a thick line between *C3* and *C5* in the *fig.5*. *PPS* updates the entry for video in *SCL* by removing *C4* from the client list is shown in *fig.5*

If it is not present in its cache, the $IM_{PSq}$ forwards the request to its parent *TR*, *VDM* at *TR* searches its database using perfect hashing to see whether it is present in any of the *PSs* in that $L_p$. If the $V_i$ is present in any of the *PSs* in that $L_p$, then the *VDM* intimates the same to $SM_{TR}$ which initiates the streaming of the $(pref-1)_{Vi}$ from that *PS*, and $(pref-2)_{Vi}$ from its cache, to the requested $PS_q$ and the same is intimated to the requested $PS_q$. Then the $IM_{TR}$ coordinate with *MMS* to download the remaining portion $(S-(pref-1)-(pref-2))_{Vi}$, and hence the request-service delay is very small

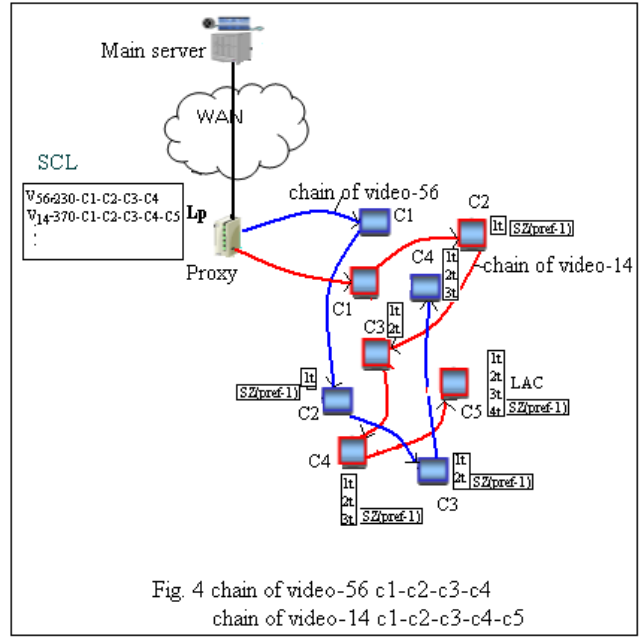

Fig. 4 chain of video-56 c1-c2-c3-c4
chain of video-14 c1-c2-c3-c4-c5

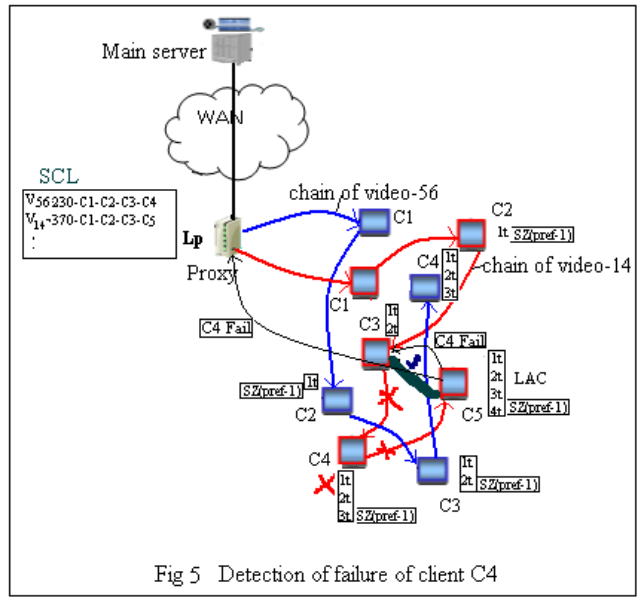

Fig 5 Detection of failure of client C4

If the $V_i$ is not present in any of the *PSs* in that $L_p$, then the $IM_{TR}$ Passes the request to the tracker of $NBR(L_p)$. Then the $VDM(NBR(L_p))$ checks its database using perfect hashing, to see whether the $V_i$ is present in any of the *PSs* of its $L_p$. If it is present, then the $SM(L_p)$ in turn initiates the streaming of $V_i$ to the requested $PS_q$ through the optimal path, and the same is intimated to the requested $PS_q$ and hence request-service delay is comparatively high but acceptable because it bypasses the downloading of the complete video from *MMS* using *MMS-PS WAN bandwidth*.

If the $V_i$ is not present in any of the *PSs* of its $NBR(L_p)$ also, then the $TR(L_p)$ modules decides to download the $V_i$ from *MMS* to $PS_q$. So the $IM_{TR}$ coordinates with *MMS* to download the $V_i$, and hence the request-service delay is very high, but the probability of downloading the complete video from *MMS* is very less as shown by our simulation results.

Whenever the sufficient buffer and bandwidth is not available in the above operation the user request is rejected

### VII. PERFORMANCE EVALUATION

In comparison with the prefix caching without chaining (PC-Chaining), we use $T=sz(pref-1)$ as the threshold value and following performance metrics in our evaluation i) Client rejection ratio $R_{rej}= \frac{Nrej}{R}$ where $N_{rej}$ is the number of requests rejected, $R$ is the total number of requests arrived ii) Average network bandwidth usage on *MS-PS* path $BW= \sum_{i=1}^{Q} bwi$ iii) Main Server load $S_L$ iv) Number of requests served from $L_p$.

**Client rejection ratio & Video hit ratio:** Fig.8 shows the client rejection ratio of our proposed system, which is very less. We can observe that PC+Chaining allows more clients to join the chain, because as the request arrival rate increases T $W(pref-1)$ is also increases. This keeps IS-STREAMING flag of some client in the chain true. So the request can be served immediately from the PS by adding it to the existing chain, which increases the video hit ratio as shown in fig.7 and significantly reduces the client waiting time as shown in the fig.11

**Average server-proxy bandwidth usage:** The main server may reject many requests due to its bandwidth constraints, when the clients directly request the main server. Our proposed $L_p$ architecture and a scheme of caching maximum portion of (pref-1) and (pref-2) of most frequently asked videos at PS and at TR respectively, and load sharing algorithm with PC+Chaining increases the system efficiency and reduces the average MMS-PS network bandwidth usage as shown in fig.6. and fig.9

Table.2
Simulation results

| Notation | System Parameters | Results |
|---|---|---|
| Q (Total) | Total No. of requests served at Lp | 320 |
| Q (PS) | No. of requests served with PC+Chaining from Lp | 296(93%) |
| | No. of requests served with PC-Chaining from Lp | 198(65%) |
| $BW^{S-P}$ | Mean bandwidth usage from server to Proxy | 21% |
| VHR | Average Video Hit ratio | 92% |
| Y | Client waiting time | 2sec - 3Sec |
| $S_L$ | MMS Load reduction | 70% |

**Main Server load:** As popularity based (pref-1) and (pref-2) of most of the videos are cached and streamed from $L_p$ itself, main server is contacted for least amount of video data ((S-(pref-1)-(pref-2)), and for the complete video very rarely, which has reduced the server load significantly as shown in fig.10. This in turn reduces the WAN bandwidth usage from MMS to PS as shown in the Fig. 9.

**Impact of PC+Chaining on Client waiting Time :** Our proposed approach increases the **ag**gregate storage space of $L_p$ by distributing the large number of videos across the *PSs* and *TR* of $L_p$, and hence achieves the high cache hit rate there by reducing the client waiting time significantly. For example, if 10 *PS*s within a *LPSG* managed 500 Mbytes each, total space available is 5 GB. 200 proxies of *LPSG* could store about 100 GB of movies. Generally the scalability of PC-Chaining improves because of the popularity based presence of the video at any one of the PS in $L_p$. As the size of the (Pref-1) increases, more and more number of clients can be joined to the existing chain of the video and they can be served immediately from the $L_p$ itself. This reduces the client waiting time significantly as shown in the fig.11, which also reduces the WAN bandwidth usage along server-proxy path as depicted in the fig.9.

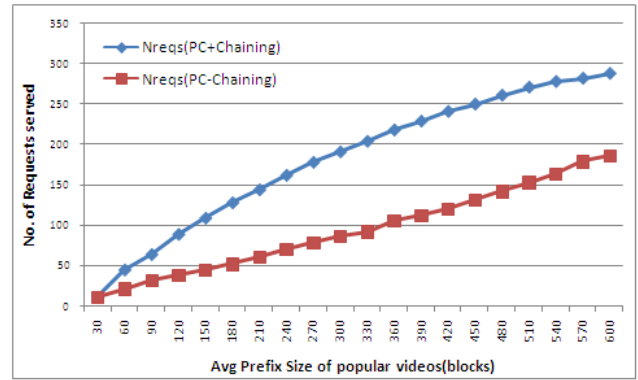

Fig.6 No. of Requests served vs Prefix Size of popular videos

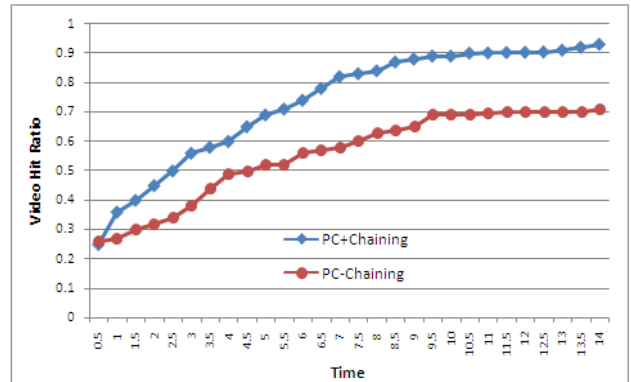

Fig. 7 Video Hit Ratio at LPSG, PC with chaining vs without chaining

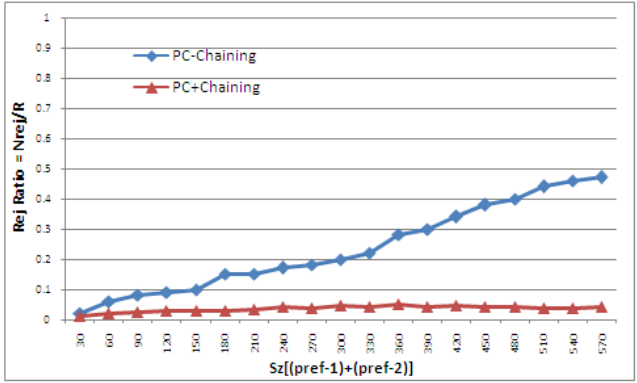

Fig.8 Reduction of client Rejection ratio as the Size of the Prifix increases (prfix caching with chaining vs without chaining)

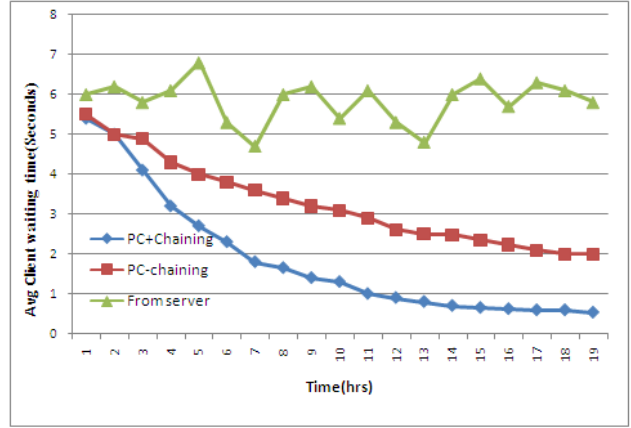

Fig.11 Average Client waiting time vs Time

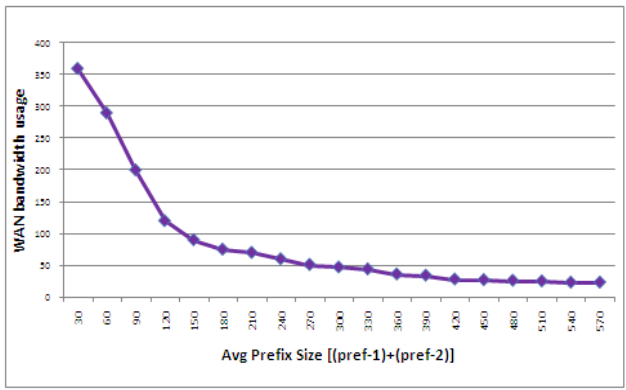

Fig.9 Avg Prefix Size [(pref-1)+(pref-2)] vs WAN bandwidth Usage

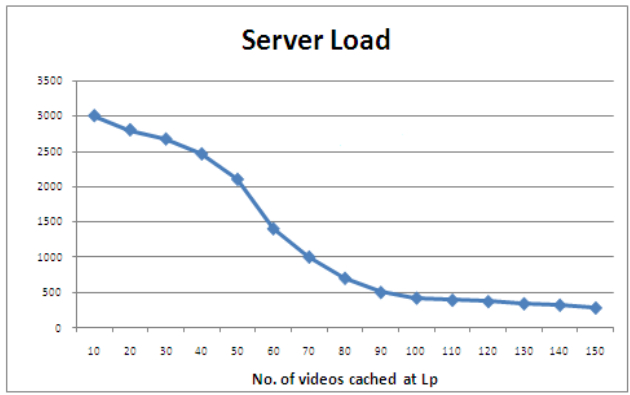

Fig.10 Reduction of MMS as the No. of videos cached at Lp increases

## VIII. CONCLUSION

In this paper we have proposed an efficient video sharing mechanism with chaining and an architecture where proxies and clients cooperate with each other to achieve reduced network bandwidth usage on the server-proxy path and client rejection ratio, by caching and streaming maximum portion of the most frequently requested videos from $L_p$. and by sharing the video data of the currently played video object with other existing clients. And the popularity based W(pref-1) allows more number of clients to get benefit from the existing chain. Our simulation results demonstrated that our proposed approach has reduced the average bandwidth usage on server-proxy path, and also the load of main server, client rejection ratio and achieves the client failure recovery successfully. The future work is being carried out to improve the performance using dynamic buffer management at PS.

AUTHORS PROFILE

**M Dakshayini** holds M.E degree in Computer science. She has one and half decades of teaching experience. she has published many papers. She is currently working as teaching faculty at Information and science and engineering department, BMS College of engineering, Bangalore, India.

**T.R. Gopalakrishnan Nair** holds M.Tech. (IISc, Bangalore) and Ph.D. degree in Computer Science. He has 3 decades experience in Computer Science and Engineering through research, industry and education. He has published several papers and holds patents in multi domains. He won the PARAM Award for technology innovation. Currently he is the Director of Research and Industry in Dayananda Sagar Institutions, Bangalore, India.